\documentclass[twocolumn,showpacs,preprintnumbers,amsmath,amssymb]{revtex4}

\usepackage{graphicx}
\usepackage{dcolumn}
\usepackage{bm}

\newcommand{\fpr}[1]{\left( #1\right)}

\newcommand{\fsq}[1]{\left[ #1\right]}
\newcommand{\fang}[1]{\langle #1\rangle}
\newcommand{\fabs}[1]{\left| #1\right|}
\newcommand{\bfx}{{\bf{x}}}

\newcommand{\WMAP}{{\it{WMAP }}}
\newcommand{\fnl}{f_{\mathrm{NL}}}
\newcommand{\side}{{\mathrm{side}}}
\begin{document}

\title{Probing Non-Gaussianity In The Cosmic Microwave Background Anisotropies: One Point Distribution Function}
\author{E. Jeong$^1$ and G. F. Smoot$^{1,2}$}
\email{ehjeong@berkeley.edu,   gfsmoot@lbl.gov}
\affiliation{$^1$Department of Physics, University of California, Berkeley, 
CA, 94720\\
$^2$Lawrence Berkeley National Laboratory, 1 Cyclotron Road, Berkeley, CA, 
94720}

\date{\today}

\begin{abstract}
We analyze \WMAP 3 year data using the one-point distribution functions to 
probe the non-Gaussianity in the Cosmic Microwave Background (CMB) Anisotropy 
data. Computer simulations are performed to determine the uncertainties of the 
results. We report the non-Gaussianity parameter $\fnl$ is constrained to 
$26<\fnl<82$ for Q-band, $12<\fnl<67$ for V-band, $7<\fnl<64$ for W-band 
and $23<\fnl<75$ for Q+V+W combined data at 95\% confidence level (CL).   
\end{abstract}

\pacs{Valid PACS appear here}
\maketitle

\section{introduction}
Non-Gaussianity is one of the most important tests of models of the inflation. 
Among the various theoretical models on the inflation, slow-roll inflation is 
currently most lively being studied. There are various predictions on the 
magnitude of non-Gaussianity based on the simple model of slow-roll 
inflation and its extensions, ranging from undetectably tiny values to large 
enough values to be detectable with currently available data 
\cite{barnabycline,battefeldeasther,calcagni,creminelli,bartolo.et.al}. On the 
other hand, observational works have claimed both detection and non-detection 
of non-Gaussianity (for reviews on recent works, see 
\cite{komatsu.et.al,Spergel.et.al,troia.et.al,creminelli.et.al,gaztanagawagg}).
Among the popular techniques for detecting non-Gaussianity are one-point 
distribution function fitting, bispectrum, trispectrum and Minkowski 
functionals. Here, we investigate the one-point distribution functions to 
probe primordial non-Gaussianity in the CMB anisotropy data. An observed CMB 
anisotropy at a 
direction ($\delta T_{obs}$) can be regarded as the superposition of three 
parts: physical fluctuation of cosmic origin ($\delta T_p$), instrumental noise 
($\delta T_n$), and foreground emissions ($T_{fg}$). Since the foreground 
templates are separately prepared, we start with foreground-removed data of 
which the CMB anisotropy can be decomposed into two uncorrelated components, 
\begin{equation}\label{eq1}
\delta T=\delta T_{obs}-T_{fg}=\delta T_{p}+\delta T_n.
\end{equation}
The primary source for the cosmic fluctuation of CMB at the large scale is 
attributed to the Sachs-Wolfe effect which is again triggered by the 
primordial curvature perturbation. The curvature perturbation $\Phi$ by 
primordial seed during the inflation is transferred to CMB anisotropy with the 
relation
\begin{equation}\label{eq2}
\frac{\delta T_p\fpr{\bfx}}{T_0}=\eta_{t}\Phi\fpr{\bfx}
\end{equation}
where $T_0=2.725$ K, the thermodynamic temperature of the CMB today, and 
$\eta_{t}$ is the radiation transfer function. For the super-horizon scale, 
we take $\eta_{t}=-1/3$ from the Sachs-Wolfe effects. At the first-order of 
perturbation, we may replace $\Phi =\Phi_g$, where $\Phi_g$ is an auxiliary 
random Gaussian field with its mean $\fang{\Phi_g}=0$ and its variance denoted 
by $\fang{\Phi_g^2}$. When the second-order perturbation is considered, it is 
conventional to prescribe the nonlinear coupling of the curvature perturbation 
as \cite{komatsuspergel} 
\begin{equation}\label{eq3}
\Phi (\bfx )\simeq\Phi_g(\bfx )+\fnl\fpr{\Phi_g^2(\bfx )-\fang{\Phi_g^2}}
\end{equation}
where $\fnl$ is the non-Gaussianity parameter. The second term in 
\eqref{eq3} is responsible for the non-Gaussianity of the primordial 
fluctuation. Then, the probability distribution function of the non-Gaussian 
field $\Phi$ can be derived as
\begin{eqnarray}
f_{\Phi}(\Phi)&=&\int f_G(\Phi_g)\delta_D
\fsq{\Phi-\Phi_g-\fnl\fpr{\Phi_g^2-\fang{\Phi_g^2}}}d\Phi_g\nonumber\\
&=&\frac{1}{\sqrt{2\pi\fang{\Phi_g^2}\fnl^2\fpr{\Phi_+-\Phi_-}^2}}\nonumber\\
& &\times\fsq{\exp\fpr{-\frac{\Phi_+^2}{2\fang{\Phi_g^2}}}
+\exp\fpr{-\frac{\Phi_-^2}{2\fang{\Phi_g^2}}}}\label{eq4}
\end{eqnarray}
where $\Phi_{\pm}$ are defined by
\begin{equation}\label{eq5}
\Phi_{\pm}=\frac{1}{2\fnl}
\fsq{-1\pm\sqrt{1+4\fnl\Phi+4\fnl^2\fang{\Phi_g^2}}}
\end{equation}
and $\Phi$ has to be limited by the reality of $\Phi_{\pm}$ as
\begin{equation}\label{eq6}
\fnl\Phi > -\frac{1}{4}-\fnl^2\fang{\Phi_g^2}.
\end{equation}
$\fang{\Phi_g^2}$ can be expressed in terms of $\eta_t$, $T_0$ and 
$\sigma_{CMB}$,
\begin{equation}\label{eq7}
\fang{\Phi_g^2}=\frac{1}{4\fnl^2}\fsq{-1+\sqrt{1+8
\fpr{\frac{\fnl\sigma_{CMB}}{\eta_t T_0}}^2}}.
\end{equation}
For a pixelized CMB anisotropy data set, the probability distribution function 
for Gaussian instrumental noise becomes
\begin{equation}\label{eq8}
f_N(\delta T_n)=\frac{1}{N_{pix}}\sum_{i=1}^{N_{pix}}
\frac{1}{\sqrt{2\pi\sigma_0^2/n_i}}
\exp\fsq{-\frac{\delta T_n^2}{2\sigma_0^2/n_i}}
\end{equation} 
where $n_i$ is the effective number of measurements at the $i_{th}$ pixel and 
$\sigma_0$ represents the dispersion of the instrumental noise per observation 
($\sigma_0$=2.1898, 3.1249, 6.5112 mK for Q, V, W-band, respectively 
\cite{Limon.et.al}). Now, it is straightforward to express the probability 
density function for $\delta T$ in an integral form,
\begin{eqnarray}
f(\delta T)&=&\int f_{\delta T_p}(\delta T_p)f_N(\delta T_n)\nonumber\\
& &\times\delta_D\fpr{\delta T-\delta T_p-\delta T_n}
d\delta T_p d\delta T_n\nonumber\\
&=&\int f_{\Phi}(\Phi )f_N(\delta T_n)\nonumber\\
& &\times\delta_D\fpr{\delta T-\eta_t T_0\Phi-\delta T_n}d\Phi d\delta T_n.
\label{eq9}
\end{eqnarray}
The probability density function derived in \eqref{eq9} explicitly contains 
the non-Gaussianity parameter $\fnl$, and it can serve as the prediction of 
one-point distribution function with a given $\fnl$ for a (ideally) 
foreground-removed CMB anisotropy data set to estimate the magnitude of 
deviation from Gaussian distribution in a quantitative manner.
\section{Application to \WMAP data}
\begin{figure*}[ht]
\begin{center}
\includegraphics[width=6.2cm,angle=0]{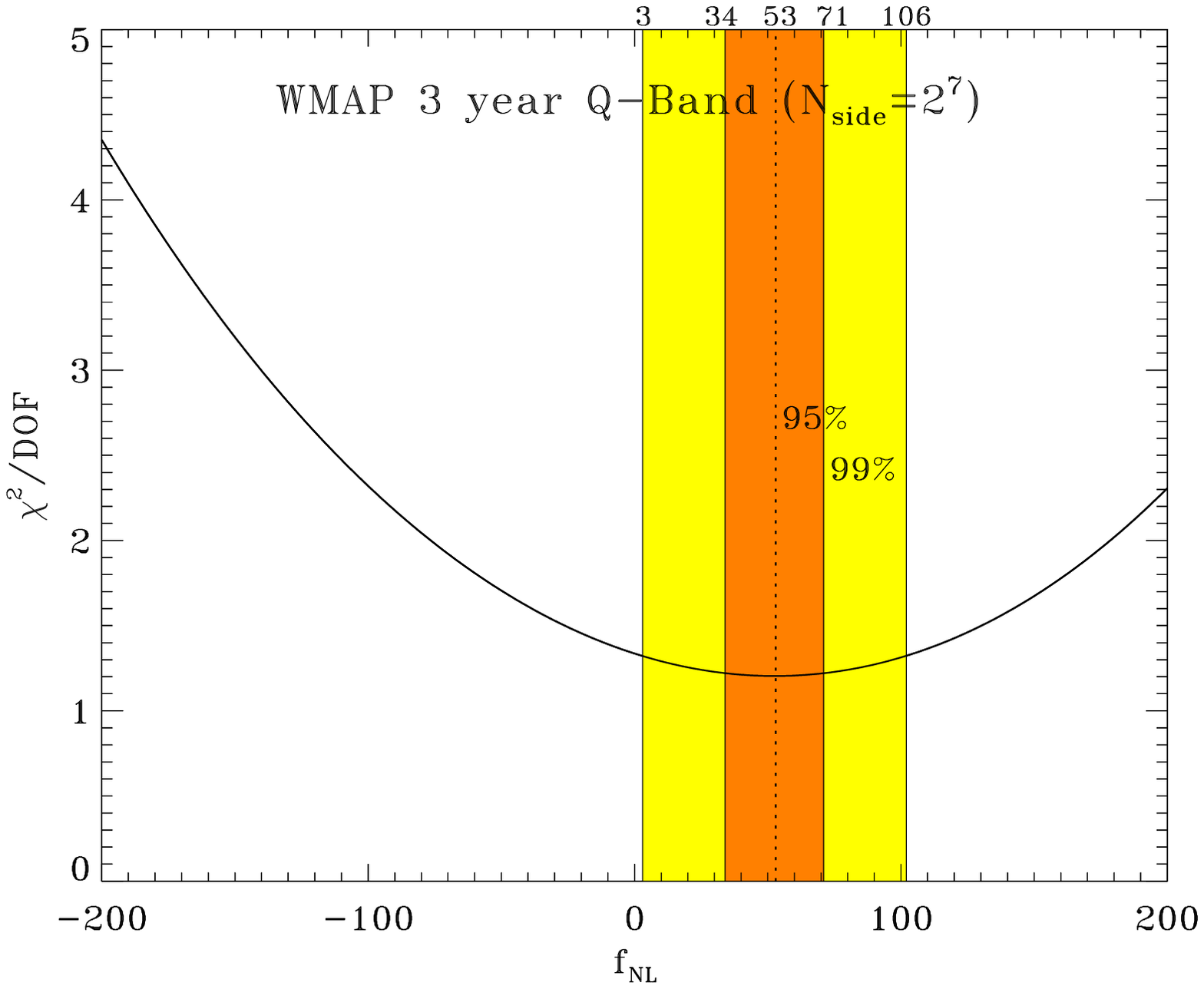}
\includegraphics[width=6.2cm,angle=0]{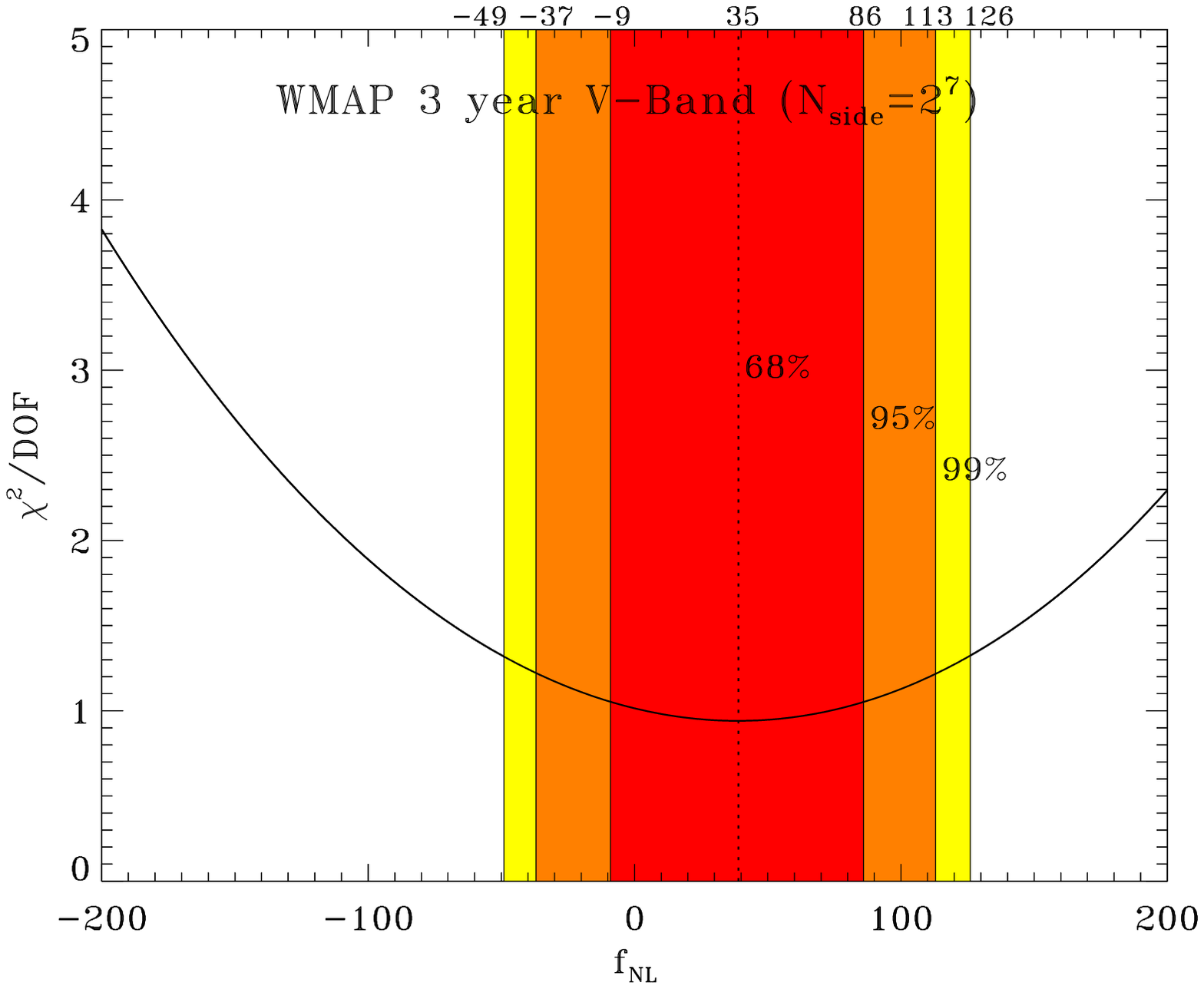}
\includegraphics[width=6.2cm,angle=0]{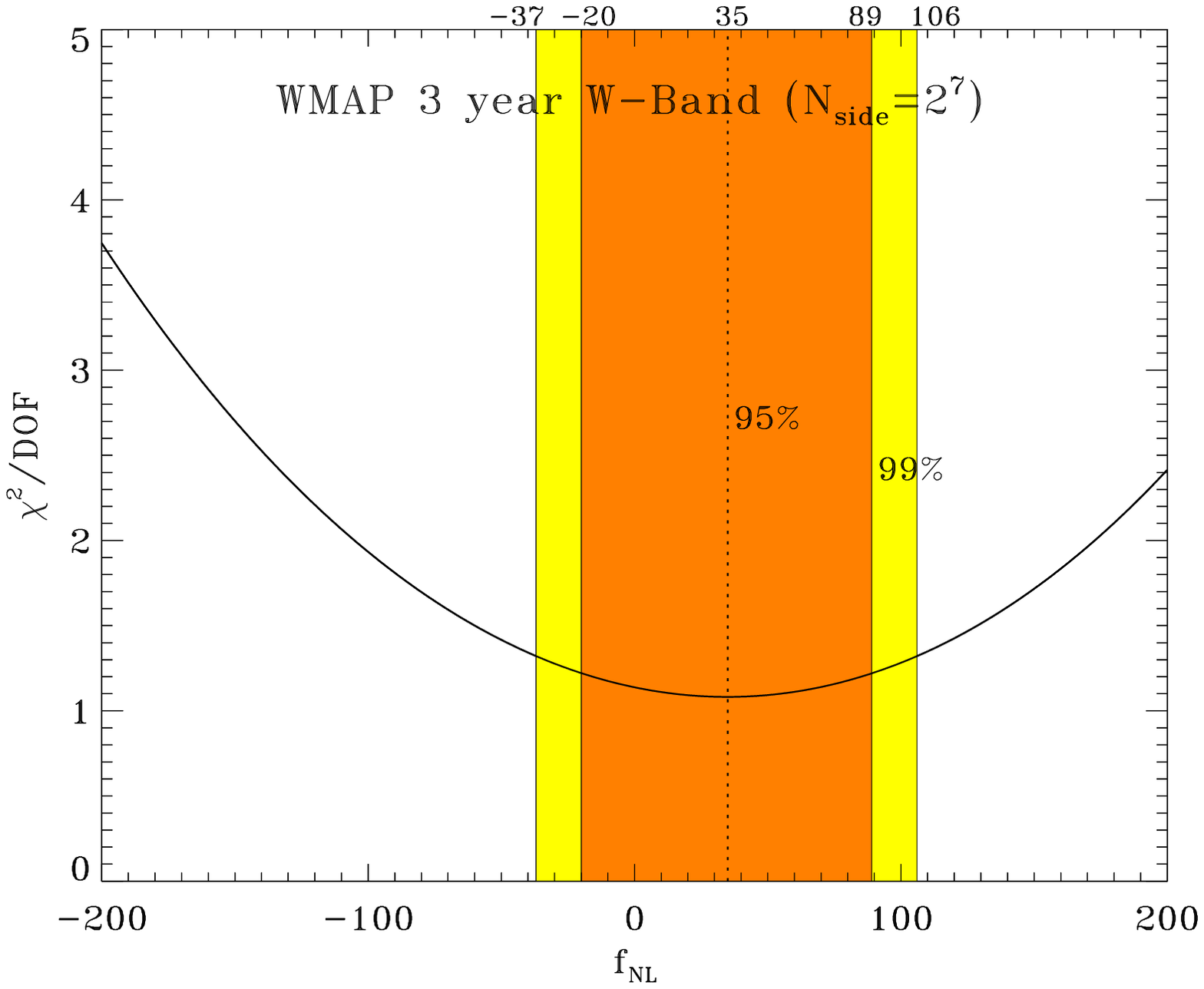}
\includegraphics[width=6.2cm,angle=0]{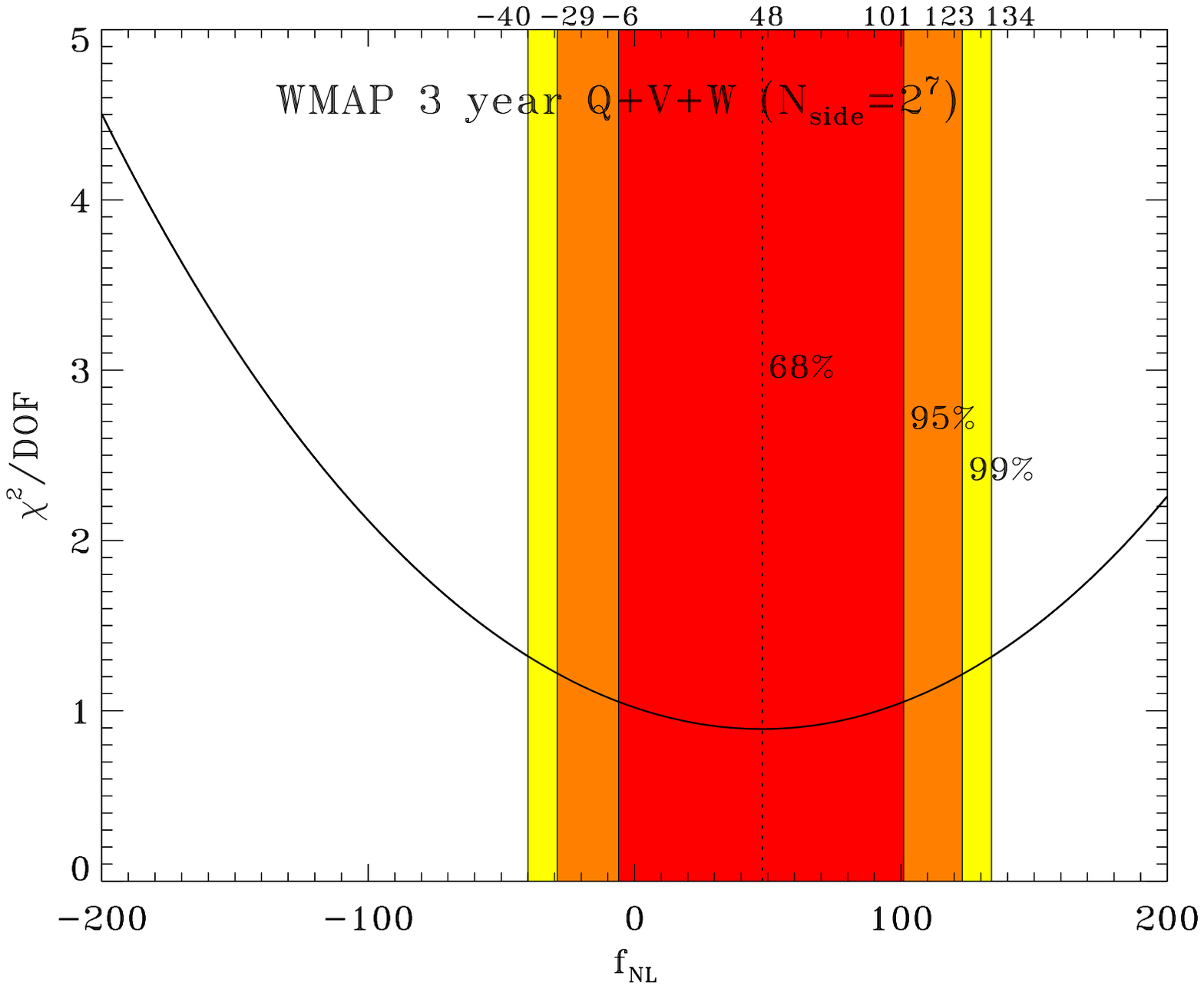}
\caption[$\fnl$ vs. $\chi^2$]
{\label{fig1}$\chi^2$-test for goodness of fit is used to find the optimal
value of the non-Gaussianity parameter. The anisotropy maps are demoted to
$N_{\side}=2^7$. Top left: Q-band, top right: V-band,
bottom left: W-band, bottom right: Q+V+W combined. The color (shaded) bands
indicate the bounds on $\fnl$ in which the probability that the statistic 
$\chi^2$ is smaller than that of the observed with respect to the prediction 
curve with $\fnl$.}
\end{center}
\end{figure*}

We use the three channels of \WMAP 3 year CMB anisotropy data sets (Q (33GHz)-, 
V (61GHz)-, W (94GHz)-band) which contain dominant signal over contaminations 
to investigate the non-Gaussianity of the CMB anisotropy data. To remove the 
foreground emissions, the Maximum Entropy Method (MEM) maps of the 
synchrotron, free-free and thermal dust are 
used\footnote[1]{http://lambda.gsfc.nasa.gov/}. The Kp0-mask is 
applied to the sky maps to remove the intense Galactic emissions and scattered 
bright point sources, which leaves 76.5\% of the sky. We also prepare a 
combined map (Q+V+W) by taking a weighted sum for a pixel temperature,
\begin{equation}\label{eq10}
\delta T\fpr{\bfx}=\frac{\sum_i\delta T_i\fpr{\bfx}n_i\fpr{\bfx}/\sigma_{0i}^2}
{\sum_in_i\fpr{\bfx}/\sigma_{0i}^2},\quad i={\mathrm{Q,\: V,\: W}}
\end{equation}
where $n_i\fpr{\bfx}$ is the effective number of measurements at the pixelized 
position $\bfx$ and $\sigma_{0i}$ is the dispersion of the instrumental noise 
of the $i_{\mathrm{th}}$ channel. We can trace the effective variance of 
the instrumental noise as a result of weighted sum defined in \eqref{eq10} as
\begin{equation}\label{eq11}
\sigma^2\fpr{\bfx}=\fsq{\sum_in_i\fpr{\bfx}/\sigma_{0i}^2}^{-1}
,\quad i={\mathrm{Q,\: V,\: W}}.
\end{equation}
The sky map data are degraded from $N_{\mathrm{side}}=512$ 
($6.87^{\prime}$ pixel) to $N_{\mathrm{side}}=128$ ($27.48^{\prime}$ pixel) 
where the number of pixels in a full sky map is given 
by $12\times N_{\mathrm{side}}^2$. The purpose of demotion of the resolution is 
to suppress the small scale fluctuation which is dominated by the instrumental 
noise. We perform the $\chi^2$-test for the goodness of fit for the probability 
density function given in \eqref{eq9} as a prediction to the observed 
probability density function which is directly 
calculated from the \WMAP data. Figure \ref{fig1} and Table \ref{t1} show the 
results of $\chi^2$ fitting of \WMAP data sets with varying $\fnl$ as a free 
parameter. All data sets are best fitted at positive $\fnl$ 
(dubbed $\fnl^{\fpr{\mathrm{opt}}}$) which are consistent with one another as 
well as the results with previous work \cite{Spergel.et.al} within the 
statistical errors.
\begin{table}[ht]
\begin{center}
\begin{tabular}{c|r@{$<\!\fnl\!<$}l|r@{$<\!\fnl\!<$}l|r@{$<\!\fnl\!<$}l
|r@{$<\!\fnl\!<$}l}
\hline\hline
Map&\multicolumn{2}{c|}{Q-band}&\multicolumn{2}{c|}{V-band}
&\multicolumn{2}{c|}{W-band}&\multicolumn{2}{c}{Q+V+W}\\
\hline
$\fnl^{\fpr{\mathrm{opt}}}$&\multicolumn{2}{c|}{53}&\multicolumn{2}{c|}{39}
&\multicolumn{2}{c|}{35}&\multicolumn{2}{c}{48}\\
\hline
68\%&\multicolumn{2}{c|}{N/A}&-9&86&\multicolumn{2}{c|}{N/A}&-6&101\\
95\%&34&71&-37&113&-20&89&-29&123\\
99\%&3&102&-49&126&-37&106&-40&134\\
\hline
DOF&\multicolumn{2}{c|}{119}&\multicolumn{2}{c|}{119}
&\multicolumn{2}{c|}{119}&\multicolumn{2}{c}{119}\\
\hline\hline
\end{tabular}
\end{center}
\caption{\label{t1}The results of $\chi^2$-tests for Goodness-of-fit for \WMAP 
3 year data. The percentages on the first column represent the tail 
probabilities at which the statistic $\chi^2$ would be smaller than the 
observed. DOF on the bottom row stands for {\it{Degrees of freedom}}.}
\end{table}

\section{Simulation And Statistical Uncertainties}
\begin{figure*}[ht]
\begin{center}
\includegraphics[width=6.2cm,angle=0]{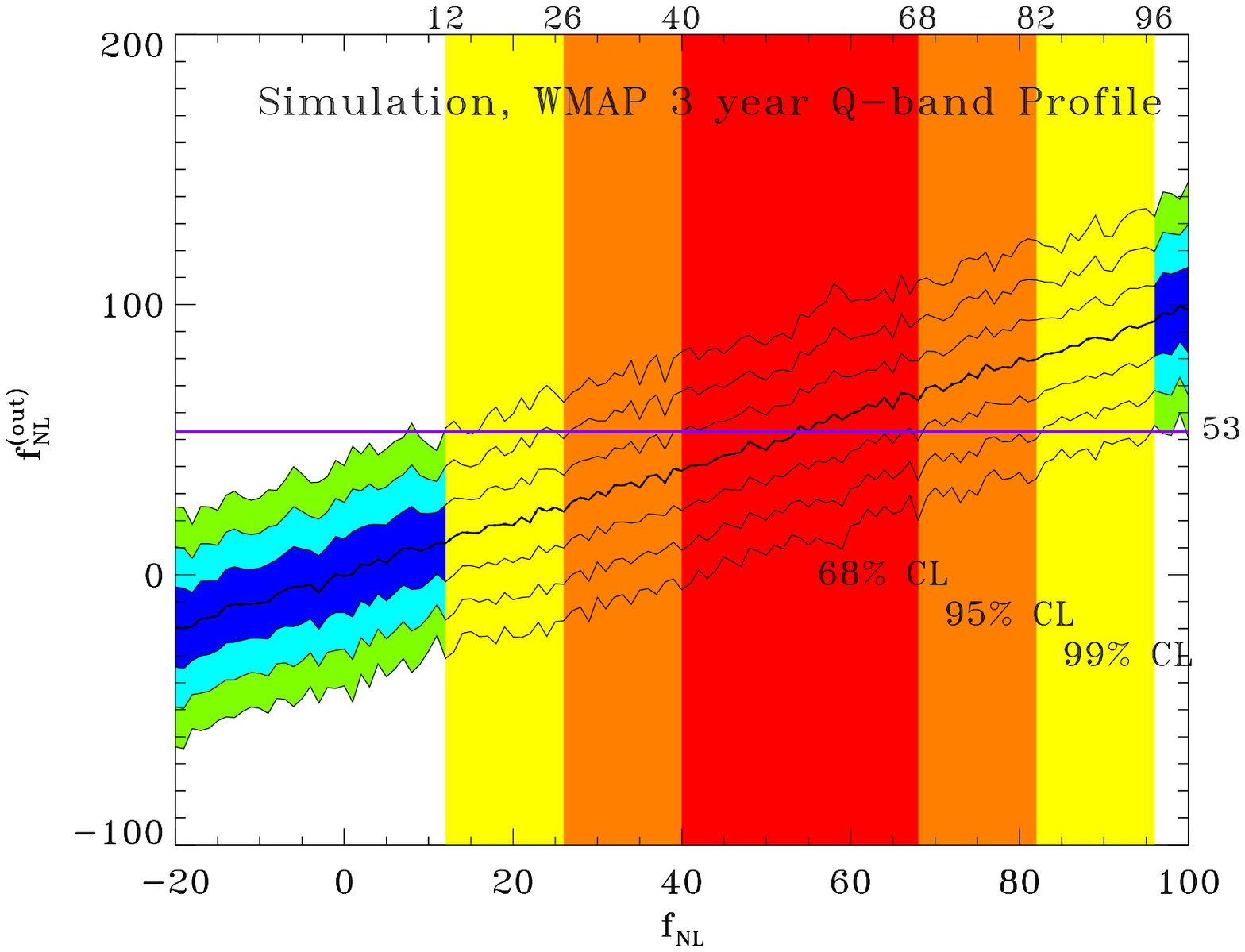}
\includegraphics[width=6.2cm,angle=0]{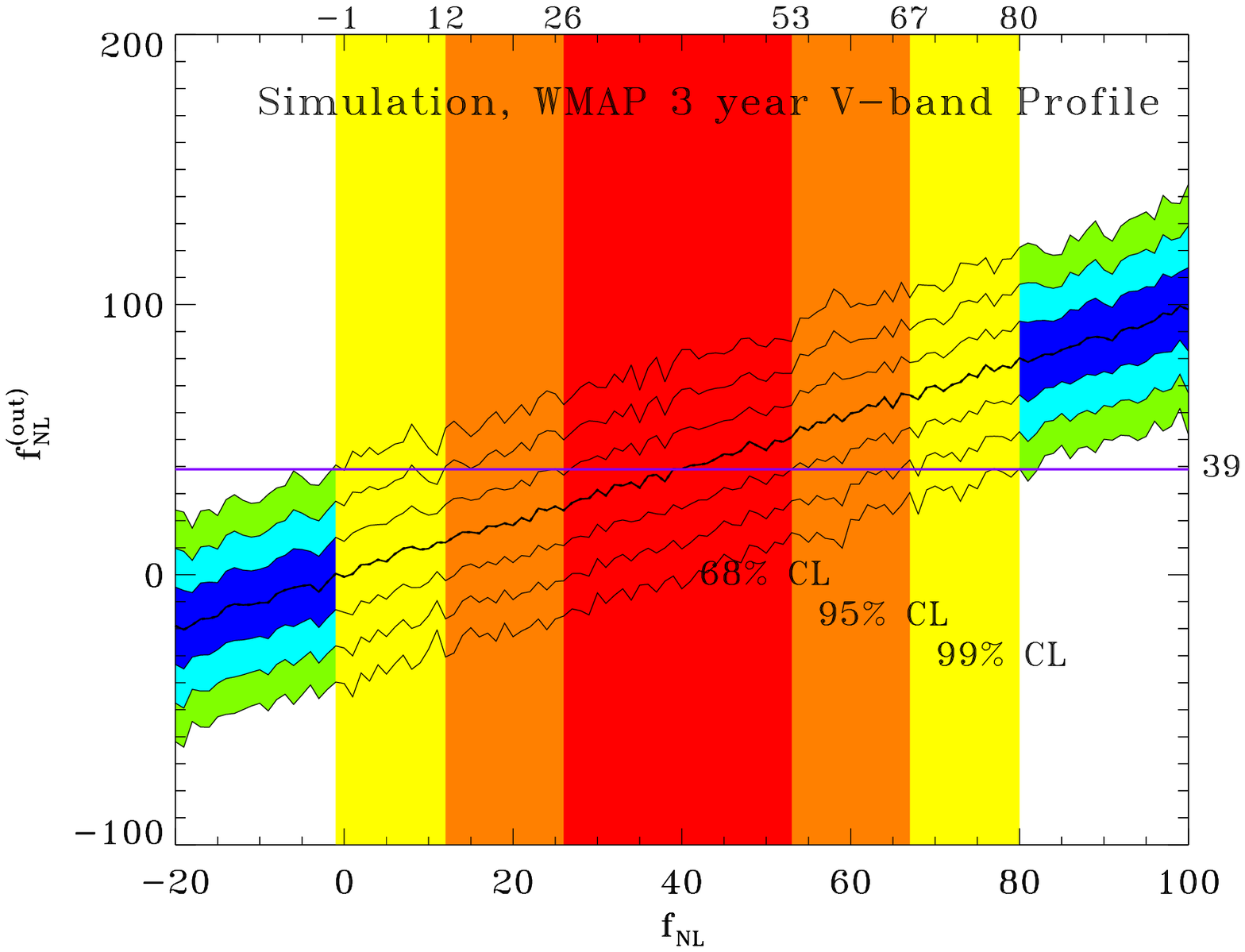}
\includegraphics[width=6.2cm,angle=0]{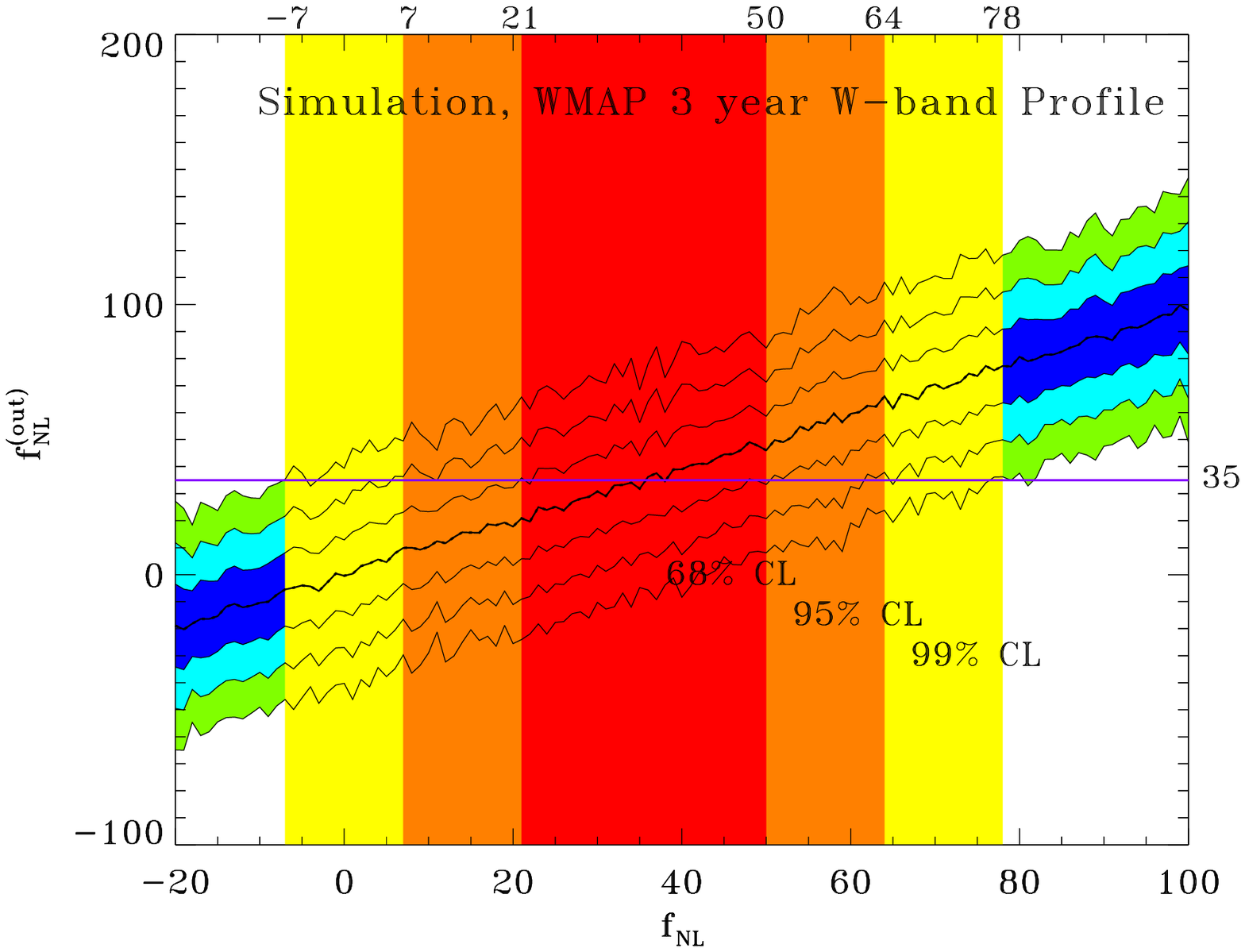}
\includegraphics[width=6.2cm,angle=0]{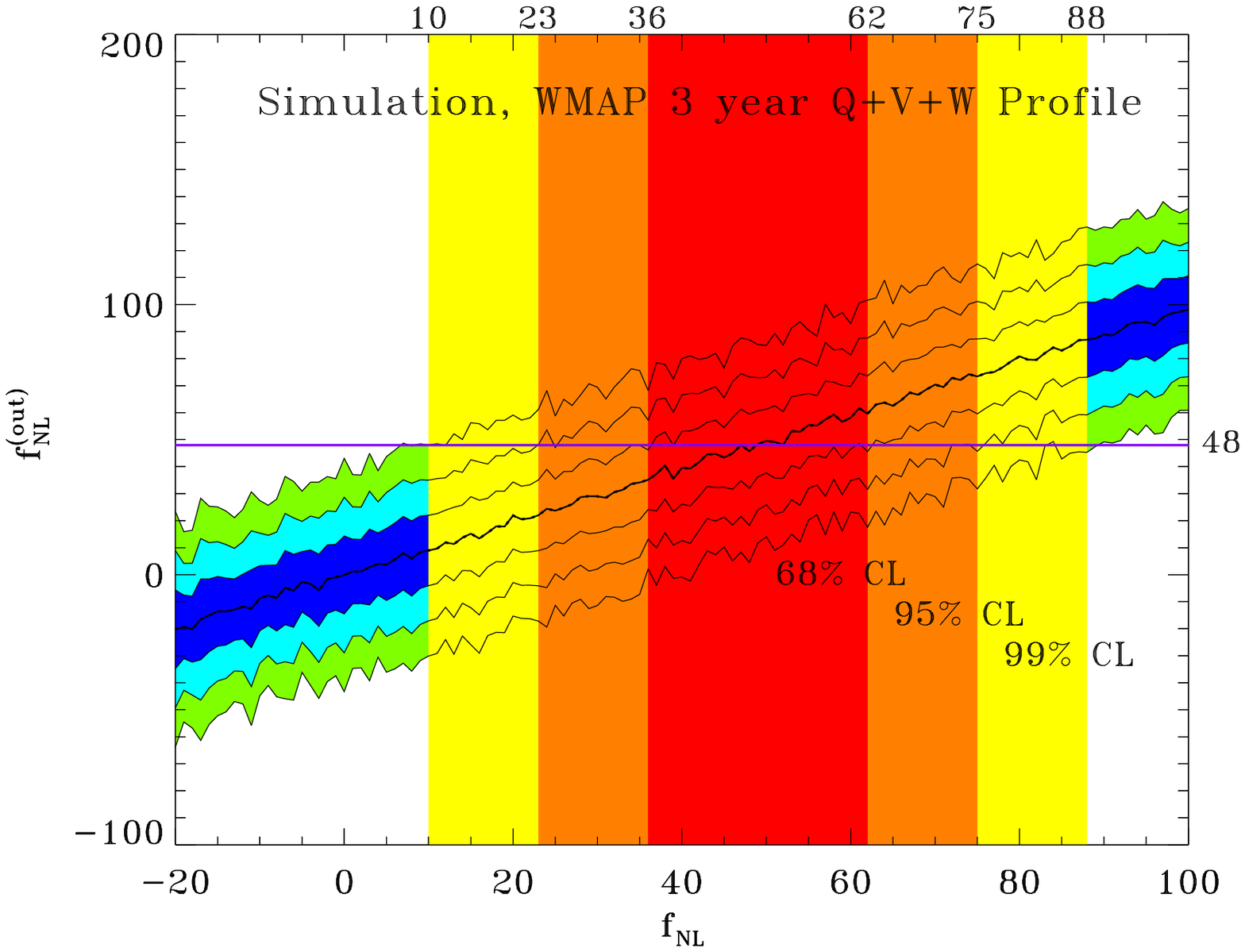}
\caption[Simulation]
{\label{fig2}Error bands found through the simulations with the same
profiles with \WMAP 3 year data (blue-green slanted bands). Here the 
``same profile'' means that the simulated data set has exactly same 
instrumental noise structure and $\sigma_{CMB}$ as a real \WMAP data set has. 
The vertical bands indicate the bounds for $\fnl^{\fpr{\mathrm{opt}}}$ found 
from the analysis of the \WMAP data.  Top left: Q-band, top right: V-band, 
bottom left: W-band, bottom right: Q+V+W. On the vertical axis, 
$\fnl^{\fpr{\mathrm{out}}}$ is the measured $\fnl^{\fpr{\mathrm{opt}}}$ from 
the simulated data.}
\end{center}
\end{figure*}

As is shown in Figure \ref{fig1}, \WMAP data fit well with finite range of the 
non-Gaussianity parameter $\fnl$. We pick $\fnl^{\fpr{\mathrm{opt}}}$ as the 
representative magnitude of non-Gaussianity for a data set, and carry out the 
Monte-Carlo simulations to test the pertinence of $\fnl^{\fpr{\mathrm{opt}}}$ 
as a proper measure of non-Gaussianity for a data set in a quantitative manner. 
A simulated data set is prepared as follows: First, a Gaussian field $\Phi_g$ 
with its variance equal to \eqref{eq7} is generated and we use it to generate 
$\Phi$-field of which the deviation from Gaussianity is denoted by $\fnl$. 
Second, we prepare a noise map in which each pixel contains a random value 
picked from a normal distribution with variance $\sigma_0^2/n_i$ as defined in 
\eqref{eq8} and add this to $\Phi$-field. A simulated map generated in this 
way has the same noise structure and the dispersion of physical fluctuation 
($\sigma_{CMB}$) as a real data set. Since a data set contains nontrivial 
instrumental noise and the number of pixels is finite, the returned magnitude 
of non-Gaussianity ($\fnl^{\fpr{\mathrm{opt}}}$) would have some uncertainty. 
For a given value of $\fnl$, we repeat the simulation and find that the 
algorithm returned an unbiased, normal distribution of 
$\fnl^{\fpr{\mathrm{opt}}}$ which is centered at the input value of $\fnl$. 
Thus, from the results of the simulations, we are able to set the bounds on 
$\fnl$ for the real data. The simulation results and error bands are plotted 
in Figure \ref{fig2} and the deduced uncertainties for $\fnl$ are summarized 
in Table \ref{t2}. It is very remarkable that this analysis strongly disfavors 
the null hypothesis ($\fnl=0$) and all the data sets show consistent results 
within the statistical errors. The results are also consistent with the works 
by the \WMAP team ($-54<\fnl<114$ at 95\% CL from bispectrum 
\cite{Spergel.et.al}) but with much tighter limits and more importantly, it 
excludes $\fnl=0$ at 95\% CL.       

\begin{table}[ht]
\begin{center}
\begin{tabular}{c|r@{$<\fnl<$}l|r@{$<\fnl<$}l|r@{$<\fnl<$}l
|r@{$<\fnl<$}l}
\hline\hline
Map&\multicolumn{2}{c|}{Q-band}&\multicolumn{2}{c|}{V-band}
&\multicolumn{2}{c|}{W-band}&\multicolumn{2}{c}{Q+V+W}\\
\hline
$N_{\mathrm{sim}}$\footnote[2]{Number of simulations for each input $\fnl$.}&\multicolumn{2}{c|}{100}&\multicolumn{2}{c|}{100}
&\multicolumn{2}{c|}{100}&\multicolumn{2}{c}{100}\\
\hline
68\%&40&68&26&53&21&50&36&62\\
95\%&26&82&12&67&7&64&23&75\\
99\%&12&96&-1&80&-7&78&9&88\\
\hline\hline
\end{tabular}
\end{center}
\caption{\label{t2}Summary of results from Simulations with \WMAP 3 year data 
profiles and the bounds at three confidence levels from the simulations.}
\end{table}

\section{Conclusion}
We developed an algorithm that uses the one-point distribution function
to investigate the non-Gaussianity of CMB anisotropy data, and applied
it to \WMAP 3 year data. We found that the null result ($\fnl$=0) is
manifestly excluded at 95\% CL. The estimated magnitude of non-Gaussianity 
parameter is $23<\fnl<75$ at 95\% CL and $9<\fnl<88$ at 99\% CL for the 
(Q+V+W)-combined map. Since the quadratic term in \eqref{eq3} takes a generic 
form of Taylor series for a perturbative expansion, it is a good 
possibility that the observed non-Gaussianity in this work is a combined 
effects of various physical processes, while the primordial seeds are very 
likely to be the leading one. There are two premises we have taken in 
developing the algorithm, which, provided they are not precise enough, could 
cause non-Gaussianity of not cosmic but systematic origin: (1) the probability 
distribution function of the instrumental noise for each pixel is centered at 
zero, and (2) the foreground emissions are removed efficiently enough in the 
foreground-removed maps. The first condition can be broken when the thermal 
and radiation environments of the \WMAP satellite in its orbit are taken into 
account, while the \WMAP team assessed they are insufficient to influence the 
science data \cite{Limon.et.al}. So, we tested the effects of the alternative 
noise distributions with a random mean in each of the Gaussian 
distribution in \eqref{eq8} and the algorithm was not misled to show 
non-Gaussianity within the statistical error. It is difficult to directly 
estimate how much residual foreground emissions after foreground subtraction 
would affect the one-point distribution function. We solely rely on the 
quality of foreground templates and it is remarkably successful, showing that 
the observed total Galactic emission matches the model to less than 1\% 
\cite{bennett.et.al2,Hinshaw.et.al}. We also analyzed simulated maps which 
are (Gaussian map + foreground templates), and all the templates for Q, V and 
W-channel showed negative values of the non-Gaussianity parameter with 
$\fabs{\fnl}\sim\mathcal{O}\fpr{10^1}$ at the resolution 
$N_{\mathrm{side}}=512$.                

\begin{acknowledgments}
We would like to thank Dr. Sara Ricciardi and Dr. Oliver Zahn for valuable 
discussion and comments on the foreground emission and other topics.
Computer simulation and data analysis with \WMAP data set were done using the
HEALPix\cite{Gorski.et.al}. This work was supported by LBNL and the Department 
of Physics at University of California, Berkeley.
\end{acknowledgments}
\bibliography{ms}
\end{document}